\documentclass[12pt]{article}

\newcommand{\cov}{{\rm {cov}}}

\newcommand{\ch}{{\cal H}}
\newcommand{\D}{{\cal D}}
\newcommand{\J}{{\cal J}}
\newcommand{\cl}{{\cal L}}
\newcommand{\V}{{\cal V}}
\newcommand{\cs}{{\cal S}}

\newcommand{\bh}{{\bf H}}
\newcommand{\br}{{\bf R}}
\newcommand{\Tr}{{{\rm Tr}}}
\newcommand{\f}{f_{\rm{vacuum}}}
\newcommand{\N}{N_{\rm{QFT}}}
\newcommand{\lm}{L_2(\cs^\prime(\br^3, \mu))}

\usepackage{latexsym,amsfonts,amsmath,amsthm,amssymb,graphicx}

\title{On the problem of hidden variables for quantum field theory}
\author{Andrei Khrennikov\\
School of Mathematics and Systems Engineering\\
University of V\"axj\"o, S-35195, Sweden}

\begin{document}
\maketitle

\abstract{We show that QFT (as well as QM) is not a complete
physical theory. We constructed a classical statistical model
inducing quantum field averages. The phase space consists of square
integrable functions, $f(\phi),$ of the classical bosonic field,
$\phi(x).$  We call our model prequantum classical statistical
field-functional theory -- PCSFFT. The correspondence between
classical averages given by PCSFFT and quantum field averages given
by QFT is asymptotic. The QFT-average gives the main term in the
expansion of the PCSFFT-average with respect to the small parameter
$\alpha$ -- dispersion of fluctuations of ``vacuum field
functionals.'' The Scr\"odinger equation of QFT is obtained as the
Hamilton equation for functionals, $F(f),$ of classical field
functions, $f(\phi).$ The main experimental prediction of PCSFFT is
that QFT gives only approximative statistical predictions that might
be violated in future experiments.}


\section{Historical introduction on the problem of completeness of quantum theories}

The problem of hidden variables is closely related to the problem of
completeness of quantum mechanics that was discussed in the paper of
Einstein, Podolsky, Rosen [1] (see also Bohr's reply to Einstein in
[2]). We note that the views of Einstein and Bohr were in the
process of the permanent evolution, see, e.g., [3] for comments.
However, it is important to remark that A. Einstein was always sure
that quantum mechanics is not complete. And this was in spite of  so
called ``NO-GO'' theorems (e.g., von Neumann's theorem [4]).

A. Einstein did not believe that the wave function provides the
complete description of a quantum system. In particular, he was one
of the founders of the so called ensemble interpretation of the wave
function, see also L. Ballentine [5]. By this interpretation the so
called pure quantum state $\psi$ is not pure at all. It describes
not the state of an individual quantum system, but statistical
properties of a huge ensemble $S_\psi$ of quantum systems. Another
important remark is that investigations of A. Einstein in the late
part of his life were concentrated on  finding a pure {\it field
model} of physical reality, including quantum reality, see, e.g.,
[1].

We note to that even the last part of Schr\"odinger's life was
characterized by comeback  to creation of purely field foundation of
quantum mechanics [6], [7]. But, in contrast to Einstein,
Schr\"odinger's attitude was toward quantum field theory (Einstein
was more interested in classical field theory).

Since typically N. Bohr did not express his views clearly enough, it
is not completely clear how he understood "completeness of quantum
mechanics" [8] (see also A. Plotnitsky for detail [9], [10]). My
personal impression of Bohr's writings is that he considered
completeness with respect to physical phase space $\Omega_{\rm
phys}=\br^3 \times \br^3.$  N. Bohr was sure that it is impossible
to provide a finer description of a quantum system based on $\Omega$
than given by the $\psi$-function. However, I am not sure that he
would claim that it would be impossible to do this on the basis of a
more general model of phase space. In any event in his
correspondence with W. Heisenberg he always discussed impossibility
to provide a detailed description of quantum phenomena by using
classical coordinates and momenta [11].

This long historical introduction was presented to convince the
reader that there  still exists a possibility (in spite of a rather
common opinion) to create a model with hidden variables which would
reproduce statistical predictions of quantum mechanics. Such a model
was presented in [12]-[14]. This is a classical statistical
mechanics with phase-space $\Omega=H \times H,$ where $H$ is a real
Hilbert space. We emphasize that our phase space $\Omega$ is
different from the conventional phase space $\Omega_{\rm phys}=\br^3
\times \br^3,$ cf. with the previous discussion on  views of Bohr
and Heisenberg. It is extremely important to remark that the
conventional quantum mechanics we obtain through a very special
choice of $H$, namely $H=L_2 (\br^3),$ the space of square
integrable functions $\psi: \br^3 \to \br.$ Thus quantum mechanics
can be reproduced on the basis of classical statistical mechanics on
phase space:
\begin{equation}
\label{PS} \Omega= L_2 (\br^3) \times L_2 (\br^3).
\end{equation}
This is the space of classical vector fields, $\psi(x)=(q(x),
p(x)).$ Here the field $q(x)$
 plays the role of the (infinite-dimensional) coordinate and the field $p(x)$
plays the role of momentum.

Thus our classical field model for quantum mechanics can be
considered as the ``Einstein-Schr\"odinger dream'' (at least late
Einstein and early Schr\"odinger). The most important deviation from
the traditional ideas on  a pre-quantum classical statistical model
is that in our approach a pre-quantum model does not reproduce
precisely quantum averages $<A>_D$ (where $A$ is a quantum
observable represented by a self-adjoint operator and $D$ is a
statistical state represented by a density operator).

Quantum mechanics is a statistical approximation of pre-quantum
classical statistical field theory (PCSFT). There is a small
parameter of the model $\alpha$. Where $\alpha \to 0,$ PCSFT is
reduced to quantum mechanics. We recall that, when $h \to 0,$
quantum mechanics is reduced to ordinary classical statistical
mechanics on phase space $\Omega_{\rm phys.}$ In [12]-[14] I
identified small parameters $\alpha$ and $h.$ It seems that it was
not correct. In [15] I proposed to distinguish parameters $\alpha$
and $h.$ The parameter $\alpha$ is small in quantum mechanics, but
the Planck constant $h$ can be chosen as equal to 1 (for the Planck
system of units).

As far as I know, in quantum field theory the problem of hidden
variables was never discussed, see e.g., [16], [17]. Roughly
speaking it was meaningless to study this problem for quantum field
theory, since even for quantum mechanics there were proved various
NO-GO theorems. It was commonly believed that {\it quantum field
theory is a complete theory}. The wave function $f(\psi)$ given by
the formalism of second quantization provides the complete
description of the quantum field. However, after the publication of
papers [12]-[15] on the asymptotic solution of the problem of hidden
variables in quantum mechanics it became clear that it is not
meaningless to consider the problem of hidden variables for quantum
field theory. In particular, the postulate on completeness of
quantum field theory can be questioned.

In this paper we apply the method of {\it asymptotic dequantization}
developed in  [12]-[15] for quantum mechanics to quantum field
theory. We show that (as well as quantum mechanics) quantum field
theory can be considered as a statistical approximation of classical
statistical mechanics for a specially chosen phase space $\Omega.$
Here $\Omega$ consists of functionals $f(\phi)$ of classical fields
$\phi.$ Classical physical variables are given by functionals of
such functionals: $f \to F(f).$ Classical statistical states are
given by Gaussian ensembles of functional $f(\phi).$ In this paper
we restrict our considerations to the case of scalar boson field
$\phi(x).$ The same procedure of asymptotic dequantization can be
applied to other fields, but it needs a lot of technical efforts.

We  also remark that our investigations on asymptotic dequantization
are closely related to so called {\it contextual probabilistic
approach to quantum mechanics}, see also [18] (cf. with conditional
probabilistic approach of G. Mackey, L. Accardi,  L. Ballentine, E.
Beltrametti, W. De Muynck, S. Gudder, [19]--[23]). We found a
natural realization of the general contextual probabilistic model by
representing contexts by Gaussian ensembles of classical fields (for
quantum mechanics) or field functionals (for quantum field theory).
So called {\it prespace} [18] - space preceding both quantum
noncommutative space (given by the Heisenberg algebra) and classical
phase space $\Omega_{\rm{phys}} = \br^3 \times \br^3$ - is given by
infinite-dimensional phase space $\Omega= H \times H.$

In our model the phase space of the classical prequantum field model
is given by $\Omega= L_2(\cs^\prime(\br^3), \mu) \times
L_2(\cs^\prime(\br^3), \mu),$ where $\cs^\prime(\br^3)$ is the space
of Schwartz distributions, and $\mu$ is the Gaussian measure on
$\cs^\prime(\br^3)$ corresponding to the free boson field [].
Statistical states are represented by Gaussian measures on $\Omega.$
They describe ensembles of functionals $f(\phi)$ of classical fields
$\phi \in \cs^\prime(\br^3)$. Physical variables are given by
functionals $F(f(\cdot))$ of field functionals $f:\cs^\prime(\br^3)
\to \br.$ Quantum field operators $A$ are obtained as second
derivatives of such functionals $F$ at the zero point: $F \to
A=\frac{F^{\prime\prime}(0)}{2}.$

In our approach classical averages are not equal to quantum field
averages. There is only an asymptotic relation between the classical
average and the quantum field average. Thus the conventional quantum
field theory gives only the first order approximation of the
prequantum classical statistical model. Our prequantum field model
contains a small parameter $\alpha \to 0$. In fact, we consider a
one parameter family $M^\alpha$ of classical statistical models. QFT
is obtained as the $\lim_{\alpha \to 0}$ of $M^\alpha:$
\begin{equation}
\label{QF1} \lim_{\alpha \to 0} M^\alpha= \N,
\end{equation}
where $\N$ is the conventional quantum field model. We point out
that the problem of the classical limit of QFT has been discussed
both on physical and mathematical levels of rigorousness, see, e.g.,
[24]. The Planck constant $h$ was considered as a small parameter:
$\N \equiv \N^h, h \to 0.$ It was shown (see, e.g., [25], [26] for
the rigorous mathematical considerations) that:
\begin{equation}
\label{QFT2} \lim_{h \to 0} \N^h=M_{\rm{cl. inf.}},
\end{equation}
where $M_{\rm{cl. inf.}}$ is the classical statistical model with
the infinite dimensional phase space.

However, we study the opposite problem: to represent the QFT-model
$\N$ as the $\lim_{\alpha \to 0}$ of classical statistical models
$M^\alpha.$ In this framework QFT is just the $\alpha \to 0$
approximation of a special classical statistical model. The latter
can be called {\it prequantum classical statistical field-functional
theory,} PCSFFT.

The small parameter $\alpha$ gives the dispersion of fluctuation of
prequantum field functionals, $f(\phi):$
\begin{equation}
\label{PC} \int_{L_2(\cs^\prime(\br^3), \mu) \times
L_2(\cs^\prime(\br^3), \mu)}\left( \int_{\cs^\prime
(\br^3)}|f(\phi)|^2 d \mu(\phi)\right) d \rho(f)=\alpha
\end{equation}
Here $f(\phi)$ is a ``classical field'' on the infinite-dimensional
configuration space $\cs^\prime(\br^3)$ -- field functional, and $
\rho$ is a Gaussian measure representing an ensemble of such field
functionals.

\section{Gaussian quantization of the scalar boson field}

Let us consider the pseudo-differential operator $a=\sqrt{- \Delta +
m^2}, m > 0.$ We pay attention that the operator $a^{-1}$ is
continuous in $\cs(\br^3).$ Thus the quadratic (positively defined)
form $b(\phi, \phi)=(a^{-1} \phi, \phi)$ is also continuous on
$\cs(\br^3).$ By the Minlos-Sazonov theorem the Gaussian measure
$\mu$ with zero mean value and the covariation operator $b_\mu={\rm
cov}\; \mu=\frac{a^{-1}}{2}$ is $\sigma$-additive on the
$\sigma$-algebra of Borel subsets of the space $\cs (\br^3).$ Let us
consider the Hilbert space $L_2(\cs^\prime (\br^3), \mu)$,
consisting of functionals $f:\cs^\prime (\br^3) \to \br$ such that
$$||f||_2^2=\int_{\cs^\prime (\br^3)} f^2 (\phi) d \mu (\phi) < \infty.$$
The basic operators of QFT, e.g., free Hamiltonian $\bh_0$ and the
operator of the number of particles ${\bf N}$, are constructed with
the aid of the procedure of the second quantization, see, e.g.,
[24]. There is a natural realization of this procedure within the
calculus of infinite-dimensional pseudo-differential operators in
$\lm$, [25], [26].

Let an operator $\lambda:\cs (\br^3) \to \cs (\br^3)$ be continuous
and let it be symmetric with respect to the scalar product in $L_2
(\br^3, d x).$ Its second quantization is defined as an operator $d
\Gamma (\lambda): L_2 (\cs^\prime (\br^3), \mu) \to L_2 (\cs^\prime
(\br^3), \mu)$ which can be defined, for example, with the aid of
its symbol:
\begin{equation}
\label{SY} d\Gamma (\lambda) (q, p)=(b_\mu \lambda p, p) +
i(q,\lambda p), \; p \in \cs(\br^3), q \in \cs^\prime (\br^3).
\end{equation}
The quantization procedure is performed through the representation
of the classical field variables, $p \equiv p(x), q \equiv q(x)$ by
the operators:
\begin{equation}
\label{OP} (q, r) \to ({\bf q}, r) f(\phi)=(\phi, r) f(\phi), r \in
S (\br^3) ;
\end{equation}
\begin{equation}
\label{OP1} (s, p) \to (s, {\bf p}) f(\phi)=\frac{1}{i} \left(s,
\frac{\delta}{\delta \phi}\right) f(\phi), s \in \cs^\prime (\br^3)
.
\end{equation}
Thus
\begin{equation}
\label{OP2} d(\lambda) ({\bf q}, {\bf p})= -(b_\mu \lambda
\frac{\delta}{\delta \phi}, \frac{\delta}{\delta \phi}) + (\phi,
\lambda \frac{\delta}{\delta \phi}).
\end{equation}
For example, for $\lambda= a= \sqrt{-\Delta + m^2}$ we get the free
field Hamiltonian:
\begin{equation}
\label{OP3} \bh_0= d \Gamma(\sqrt{- \Delta + m^2})= - \frac{1}{2}
\int_{\br^3} \frac{\delta^2}{\delta \phi^2 (x)} dx + \int_{\br^3}
\phi (x) \sqrt{- \Delta + m^2} \frac{\delta}{\delta \phi (x)} dx .
\end{equation}
If $\lambda=I$ is the unit operator, then we obtain the operator of
the number of particles:
\begin{equation}
\label{OP4} {\bf N} = d \Gamma (1)= -\frac{1}{2} \int_{\br^3}
\frac{\delta}{\delta \phi (x)} (- \Delta + m^2)^{-1/2}
\frac{\delta}{\delta \phi (x)} dx  + \int_{\br^3} \phi (x)
\frac{\delta}{\delta \phi (x)} dx.
\end{equation}
We remark that these operators are not bounded in $L_2({\cal
S}^\prime(\br^3), \mu).$ But they, of course, can be approximated by
bounded operators corresponding to approximation of the kernel of
the operator $(- \Delta + m^2)^{\pm 1/2}$ by smooth functions.
Therefore in our further considerations we restrict ourselves to the
{\it QFT-model with bounded quantum field operators.}

The QFT-model is defined as the pair:
$$
\N=(\D(\Omega_c), \cl_s (\Omega_c))
$$
where $\Omega_c = L_2^{\bf C} (\cs^\prime (\br^3), \mu)$ is the
space of square integrable with respect to the Gaussian measure
$\mu$ functionals $f: \cs^\prime (\br^3) \to {\bf C}, \D$ is the
space of density operators ($D: \Omega_c \to \Omega_c, D \geq 0, \Tr
D=1), \cl_s$ is the space of self-adjoint continuous operators ($A:
\Omega_c \to \Omega_c, A^*=A$).

\section{A classical statistical model for QFT}

We choose the phase space $\Omega,$ consisting of square integrable
field functionals, $\phi \to f(\phi).$ Thus $\Omega= Q \times P,$
where $Q=P=L_2 (\cs^\prime (\br^3), \mu).$ We consider on $\Omega$
the canonical symplectic structure given by the operator
 \[J= \left( \begin{array}{ll}
 0&1\\
-1&0
\end{array}
 \right ).
 \]
Denote by $\J$ the one parametric group with the generator J:
 \[J_\theta= \left( \begin{array}{ll}
 \cos \theta & \sin \theta\\
-\sin \theta & \cos \theta
\end{array}
 \right ), \theta \in \br
 \]
A function (in fact, functions of functionals of fields $\phi \in
\cs^\prime (\br^3)$) $F: \Omega \to \br$ is called $\J$-invariant if
\begin{equation}
\label{JJ} F(J_\theta f)= F(f), \; f \in \Omega,
\end{equation}
for any $\theta \in [0, 2 \pi).$ In our further considerations the
following simple mathematical fact will play an important role:

\medskip

{\bf Lemma 1.} {\it Let $F: \Omega \to \br$ be two times Frechet
differentiable and $\J$-invariant. Then}
\begin{equation}
\label{CR} F^{\prime\prime} (0) J=J F^{\prime\prime} (0)
\end{equation}

{\bf Corollary 1.} {\it A quadratic form $F(f)=(\bh f, f)$ is
$\J$-invariant iff $\bh J=JH.$}

\medskip

Let us denote by $\Omega_c$ the phase-space $\Omega$ endowed with
the canonical complex structure induced by the symplectic structure
on it: $$\Omega_c=Q \oplus i P \equiv L_2^C (\cs^\prime (\br^3),
\mu).$$

By Lemma 1, for any $C^2$-map $F: \Omega \to \br$ which is
$\J$-invariant, its second derivative defines the ${\bf C}$-linear
operator
$$
f^{\prime\prime} (0): \Omega_c \to \Omega_c.
$$
In particular, any quadratic $\J$-invariant form $F(f)=(\bh f, f)$
can be represented in the form $F(f)=<\bh f, f>,$ where $<\cdot ,
\cdot>$ is the canonical complex scalar product on $\Omega_c:$
$$<f, g>=\int_{\cs^\prime (\br^3)} f(\phi) \overline{g(\phi)} d\mu (\phi).$$

We denote by the symbol $\Omega^{\bf C}$ the complexification
$\Omega \oplus i \Omega$ of the phase space $\Omega:$
$$\Omega^{\bf C}=[L_2 (\cs^\prime (\br^3), \mu) \times L_2 (\cs^\prime (\br^3), \mu)] \oplus i [L_2 (\cs^\prime (\br^3), \mu) \times L_2 (\cs^\prime (\br^3), \mu)].$$
The space of classical physical variables, denoted by $\V(\Omega),$
we choose in the following way: a) $F(0)=0;$ b) $F$ can be continued
to the analytic function $F: \Omega^{\bf C} \to \bf C;$ c) $F$ is
$\J$-invariant; d) $F$ has the exponential growth on $\Omega^{\bf
C}:$ \begin{equation} \label{EG} |F(f)|\leq a_F e^{r_F||f||_2}, f
\in \Omega^{\bf C}.
\end{equation}
The following simple mathematical facts will play important roles in
our future considerations.

\medskip

{\bf Lemma 2.} {\it Let a measure $\rho$ on $\Omega$ be
$\J$-invariant. Then its covariation operator $B=\cov \rho$ commutes
with the symplectic operator $J: [B, J]=0.$}

{\bf Lemma 3.} {\it A Gaussian measure $\rho$ (with the zero mean
value) is $\J$-invariant iff $[B, J]=0.$}

{\bf Lemma 4.} {\it Let a measure $\rho$ on $\Omega$ be
$\J$-invariant. Then the "coordinate" $q(\phi)$ and the "momentum"
$p(\phi)$ give the equal contributions into its dispersion:} $$
\int_{L_2 (\cs^\prime (\br^3), \mu) \times L_2 (\cs^\prime (\br^3),
\mu)} \Big (\int_{\cs^\prime (\br^3)} q^2 (\phi) d \mu (\phi) \Big)
d \rho (q,p)$$
\begin{equation}
\label{IC}= \int_{L_2 (\cs^\prime (\br^3), \mu) \times L_2
(\cs^\prime (\br^3), \mu)} \Big (\int_{\cs(\br^3)} p^2 (\phi) d \mu
(\phi) \Big) d \rho (q,p)
\end{equation}

We choose the space of classical statistical states\footnote{ They
describe ensembles of physical systems having states belonging to
the phase space $\Omega$} -- denoted by the symbol $S_G^\alpha
(\Omega)$ -- consisting of Gaussian measures on the phase space
$\Omega$ (having the zero mean value) such that:

a) the dispersion of $\rho \in S_G^\alpha (\Omega)$ equals to
$\alpha:$
$$\sigma^2 (\rho)=\int_{L_2(\cs^\prime (\br^3), \mu)
\times L_2(\cs^\prime (\br^3), \mu)} \left( \int_{\cs^\prime(\br^3)}
(q^2(\phi) + p^2(\phi)) d \mu (\phi)\right) d \rho (q, p)=\alpha,
\alpha \to 0;$$

b) any $\rho \in S_G^\alpha (\Omega)$ is $\J$-invariant:
$$
\int_{L_2(\cs^\prime (\br^3), \mu) \times L_2(\cs^\prime (\br^3),
\mu)} f(\cos \theta q + \sin \theta p), - \sin \theta q + \cos
\theta p) d \rho (q,p)
$$
$$
= \int_{L_2(\cs^\prime (\br^3), \mu)
\times L_2(\cs^\prime (\br^3), \mu)} f(q, p) d \rho (q, p).
$$
We note that
\begin{equation}
\label{T} \sigma^2 (\rho)=\Tr B,
\end{equation}
where $B={\rm cov} \; \rho$ is the covariance operator of $\rho.$ We
also point out that $\rho \in S_G^\alpha (\Omega)$ implies that $[B,
J]=0,$ see Lemma 2, and that by Lemma 4: $$\int_\Omega ||q||_2^2 d
\rho (q, p)=\int_\Omega||\rho||_2^2 d\rho (q, p)=\frac{\alpha}{2}.$$

We shall also use the complex covariation operator of $\rho,
B^c={\rm cov}^c \; \rho$ which is given by
\begin{equation}
\label{CV} <B^c u, v>=\int<u, f><f,v> d \rho (f(\cdot)),
\end{equation}
where $f(\phi)=q(\phi) + i p(\phi).$

\medskip

{\bf Lemma 5.} {\it Let a measure $\rho$ be $\J$-invariant. Then
$B^c=2B$ (in particular, there is one-to-one correspondence between
real and complex covariation operators).}

\medskip

{\bf Lemma 6.} {\it There is one-to-one correspondence between
between  Gaussian $\J$-invariant measures and complex covariation
operators: $\rho \to B^c= {\rm cov}^c \; \rho= 2{\rm cov}\; \rho.$}

\medskip

We pay attention that by using the trace with respect to the complex
Hilbert space $\Omega_c$ we can write
\begin{equation}
\label{T1} \sigma^2(\rho)=\Tr B^c
\end{equation}.
We define one parametric family of classical statistical models:
$$M^\alpha=(S_G^\alpha(\Omega), \V(\Omega)).$$

\medskip
In the Gaussian integral $\int_\Omega F( f) d\rho(f)$ we make the
scaling: $f(\phi) = \sqrt{\alpha} f(\phi).$ By considering $\alpha$
as a small parameter, $\alpha \to 0,$ and using the Taylor expansion
of analytic functionals, $F(f),$ on the space of square integrable
field-functionals, $f(\phi),$ we obtain the following asymptotic
expansion of Gaussian integrals on the phase space, see appendix for
the detailed proof (the proof for QM presented in [1] should be
modified in that way to become mathematically correct):

{\bf Lemma 7.} {\it Let $F \in \V (\Omega)$ and let $\rho \in
S_G^\alpha(\Omega).$ Then
\begin{equation}
\label{AI} <F>_\rho\equiv \int_\Omega F (q, p) d \rho (q,
p)=\frac{\alpha}{2} \Tr D^c F^{\prime\prime} (0) + O(\alpha^2),
\alpha \to 0,
\end{equation}
where $D^c={\rm cov}^c \; \rho/\alpha$ and
$$
|O(\alpha^2)|\leq \alpha^2 K_F \int_\Omega e^{r_F||f||_2} d\rho_{D^c}(f),
$$ and $\rho_{D^c}$ is the Gaussian measure
($\sqrt{\alpha}$-scaling of $\rho$) with the  complex covariation
operator $D^c.$}

\medskip

The equality (\ref{AI}) motivates the following definition of the
asymptotic projection of the one parametric family of classical
statistical models $M_\alpha$ onto the QFT-model $\N:$
\begin{equation}
\label{P} T:\V(\Omega) \to \cl_s (\Omega_c),
T(F)=F^{\prime\prime}(0)/2
\end{equation}

Thus the classical physical variable $F:\Omega \to \br$ (functional
of functionals $f(\phi)=(q(\phi), p(\phi))$ of classical fields
$\phi \in \cs^\prime (\br^3)$) is mapped into its second derivative.
This is really a projection having the huge degeneration.
\begin{equation}
\label{P1} T: S_G^\alpha (\Omega) \to \D (\Omega_c), \rho \to
D^c=\cov^c \rho/\alpha.
\end{equation}
By Lemma 3 this map is one-to-one. One can formulate previous
considerations in the form of a theorem:

\medskip

{\bf Theorem 1.} {\it The one parametric family of classical
statistical models $M^\alpha$ provides the asymptotic
"dequantization" of QFT for the scalar bosonic field. There exists
projections given by (\ref{P}) and (\ref{P1}) of spaces of classical
physical variables and statistical states onto spaces of
self-adjoint operators (quantum field operators) and density
operators such that the asymtotic equality of classical and QFT
averages take place:}
\begin{equation}
\label{AI} <F>_\rho=\alpha <T(F)>_{T(\rho)} + O(\alpha^2), \alpha
\to 0.
\end{equation}

\medskip

Denote by the symbol $\V_{\rm quad}$ the space of quadratic forms
$F: \Omega \times \Omega \to R$ which are $\J$-invariant. Thus $F(f,
f)=(Af, f),$ where $[A, J]=0.$ Let us consider the one parametric
family of classical statistical models: $M^\alpha_{\rm
quad}=(S_G^1(\Omega), \V_{\rm quad}(\Omega)).$

{\bf Corollary 1.} {\it The family $M^\alpha_{\rm quad}$ provides
the "explicit dequantization" of QFT. Both dequantization maps,
(\ref{P}) and (\ref{P1}), are one-to-one and classical and QFT
averages coincide:}

\begin{equation}
\label{AIZ} <F>_\rho \equiv \int_\Omega (A f, f) d\rho (f)=\Tr \;
\cov^c \rho A.
\end{equation}

However, we consider the explicit dequantization given by corollary
1 as a purely mathematical construction, cf. [1], [27] for QM. The
essence of correspondence between classical and quantum worlds is
the asymptotic expansion of classical statistical averages\footnote{
Of course, one should always remember that our picture of classical
world differs crucially from the conventional one which was based on
$\br^3$ space. Classical world which was approximated by QFT is
"double infinite dimensional". Its points are functionals $f$
defined on the infinite dimensional space.}.

\section{Interpretation, structure of vacuum}

The point $\f=0 \in \Omega$ we call the {\it classical vacuum
state.} This is the field functional $\f(\phi)$ which equals to zero
for any classical bosonic field $\phi \in \cs^\prime (\br^3):
\f(\phi)= (q_{\rm{vacuum}}(\phi), p_{\rm{vacuum}}(\phi))$ and
$q_{\rm{vacuum}} \equiv 0, p_{\rm{vacuum}}\equiv 0$ on $\cs^\prime
(\br^3).$

Thus the vacuum field $\f$ is defined not on the conventional
physical space $\br^3$, but on the infinite dimensional space of
classical bosonic fields $\cs^\prime (\br^3).$ This is the crucial
departure from the conventional picture of vacuum. In any event, in
our approach ``fluctuations of vacuum'' are fluctuations of the
vacuum field functional $\f(\phi),$ cf. [28]--[34]. Such
fluctuations can be described by measures on $\Omega$ having the
very small dispersion.

Such a measure represents a random variable $f(\lambda,\phi) \in
\Omega$ (here $\lambda$ is a random parameter)\footnote{So
$f(\lambda,\phi)$ is not a random field, but a random field
functional, cf. [].} and its standard deviation gives the measure of
deviation from the vacuum field functional $\f(\phi):$
$$
D f(\lambda, \phi)=E||(\lambda, \phi)-\f(\phi)||^2_2=E \Big
(\int_{\cs^\prime(\br^3)}|f(\lambda, \phi)|^2 d\mu (\phi) \Big).
$$
Thus $\sigma(f)= \sqrt{D f(\lambda, \phi)}$ tells us how much the
random field functional $f(\lambda, \phi)$ deviates from the vacuum.
Therefore we can interpret our statistical states $\rho \in
S_G^\alpha (\Omega)$ as {\it Gaussian fluctuations of vacuum.} Here
$\alpha$ can be interpreted as {\it intensity of vacuum
fluctuations.}

Let $F$ be a classical physical variable, $F:\Omega \to \br.$ We can
consider the relative intensity:
\begin{equation}
\label{RI} F_\alpha (f)=\frac{F(f)}{\alpha} \equiv
\frac{F(q,p)}{\alpha}
\end{equation}
The basic equality of the asymptotic dequantization of QFT, see
(\ref{AI}), can be written as
\begin{equation}
\label{AI1} <F_\alpha>_\rho= <T(F)>_{T_{(\rho)}} + O(\alpha), \alpha
\to 0.
\end{equation}
Thus the QFT-average $<T(F)>_{T(\rho)}\equiv \Tr T(\rho) T(F)$ gives
us the main term in the expansion of the classical average of the
relative intensity with respect to the vacuum of fluctuations,
$$
<F_\alpha>_\rho=\int_\Omega F_\alpha (f) d\rho (f)= \frac{1}{\alpha}
\int_\Omega F(f) d \rho (f) \approx <T(F)>_{T(\rho)}.
$$

\section{Quantum field Schr\"odinger equation as Hamilton equation for field functionals}

We consider the  system of Hamilton equations  on the phase space of
field functionals $\Omega=L_2 (\cs^\prime (\br^3), \mu) \times L_2
(\cs^\prime (\br^3), \mu):$
\begin{equation}
\label{SH} \dot q=\frac{\delta \cal H}{\delta p}, \; \; \dot
p=-\frac{\delta \cal H}{\delta q}
\end{equation}
where $\ch:\Omega \to \br$ is a function of the class $C^1$ (so it
is Frechet differentiable with continuous first derivative $\nabla
\ch (q, p)= (\frac{\delta \ch}{\delta q} (q, p), \frac{\delta
\ch}{\delta p}(q, p)) \in \Omega$).

We introduce the symplectic gradient of the Hamilton function:
$$
J\nabla \ch (q, p)= (\frac{\delta \ch}{\delta p} (q, p),
-\frac{\delta \ch}{\delta q} (q, p)),
$$
and we write the system of Hamilton equations in the vector form:
\begin{equation}
\label{VF} \dot f(t, \phi)= J \nabla \ch (f(t, \phi))
\end{equation}
For example, let  $\ch(f)=$
$$
\frac{1}{4} \int_{\cs^\prime (\br^3)} \Big[ \Big(
\int_{\br^3}|\frac{\delta f(\phi(x))}{\delta \phi (x)}|^2 dx\Big) +
2 \Big(\int_{\br^3} \phi (x) \sqrt{-\Delta + \mu^2} \frac{\delta
f(\phi (x))}{\delta \phi (x)} dx \Big) \overline{f(\phi)}
$$
$$
+ \Big(\int_{\br^3} |f(\phi(x))|^2 dx \Big)^2 \Big
] d \mu (\phi).
$$
Then the Hamilton equation has the form:
 \begin{equation}
 \label{VF1}
 \dot f(t, \phi)=J \bh_0 f(t, \phi) + \int_{\br^3}|f(\phi (x))|^2 dx  f(t,
 \phi),
 \end{equation}
where the linear operator $\bh_0$ is the Hamiltonian of the free
quantum bosonic field, see (\ref{OP3}). Now let us restrict our
consideration by quadratic Hamilton functions $\ch \in \V_{\rm
quad}(\Omega).$ In this case $\bh=\ch^{\prime\prime}(0)$ commutes
with the symplectic operator $J$ and, hence, the equation (\ref{VF})
can be written in the complex form:
 \begin{equation}
 \label{VFZ}
 i \dot f(t, \phi)=\bh f(t, \phi).
 \end{equation}
 This is nothing else than the Schr\"odinger equation for QFT [16],
 [24].

\medskip

{\bf Theorem 2.} {\it For $\ch \in \V_{\rm quad}(\Omega),$ the
Hamilton equation can be written as the Schr\"odinger equation for
QFT-Hamiltonian $\bh=\ch^{\prime\prime}(0).$}

\medskip

We note that quadratic Hamilton functions describe Harmonic
oscillators in the space of field functionals $f(\phi).$ If
$\bh={\rm diag}(R, R),$ then the Hamilton function
$\ch(f)=\frac{1}{2} [(Rp, p) + (Rq, q)]$ and the Hamilton equations
have the form:
 \begin{equation}
 \label{R} \dot q(t, \phi)=R p (t, \phi), \; \;
 \dot p(t, \phi)=-R q (t, \phi) .
 \end{equation}
We can call field functionals $q(t, \phi)$ and  $p(t, \phi), \phi
\in \cs^\prime (\br^3),$ {\it mutually-inducing}: the presence of
$p(t, \phi)$ induces change of $q(t, \phi)$ and vice versa, cf. with
classical electromagnetic field $E(t, x), B(t, x), x \in \br^3.$ The
system of the Hamilton equations (\ref{R}) induces the second order
equation:
 \begin{equation}
 \label{R1}
 \ddot{q} (t, \phi) + R^2 q(t, \phi)=0 .
 \end{equation}

\medskip

{\bf Theorem 3.} {\it QFT  (for the scalar bosonic field) can be
represented as classical statistical mechanics of Gaussian ensembles
of harmonic oscillators in the space of classical field functionals
$f(\phi), \phi \in \cs^\prime (\br^3).$}

 \section{Complex representation for the Hamilton dynamics}

As usual, we introduce complex variables $$f(\phi)=q(\phi) +
ip(\phi), f^* (\phi)=q(\phi)-i(\phi), \; \; \phi \in \cs^\prime
(\br^3).$$

\medskip

{\bf Proposition 1.} {\it A map $F(q, p)$ is $\J$-invariant iff
\begin{equation}
\label{GZ} F(\lambda f, \lambda^* f^*)=F(f, f^*), |\lambda|=1.
\end{equation}}

\medskip

{\bf Proposition 2.} {\it Let $F(f,f^*)$ be analytic. Then it is
$\J$-invariant iff}
\begin{equation}
\label{GZ1} F(f, f^*)=\sum_{4=0}^\infty \frac{\delta^{2n}F}{\delta
f^n \delta f^{*n}}(0) (f, \ldots, f, f^*, \ldots, f^*)
\end{equation}

In the complex variables $f$ and $f^*$ the system of Hamilton
equations can be written as
\begin{equation}
\label{GZ2} i \dot f(t, \phi)= 2 \frac{\delta \ch}{\delta f^*} (f(t,
\phi), f^*(t, \phi)),
\end{equation}
cf. [16], [17], [24], [26].

\section{Appendix: Proof of lemma 7 on asymptotic expansion of
Gaussian functional integrals}

In the Gaussian integral $\int_\Omega F( f) d\rho(f)$ we make the
scaling:
\begin{equation}
\label{LHT}  f(\phi) = \sqrt{\alpha} f(\phi).
\end{equation} We obtain:
\begin{equation}
\label{ANN1} <F>_{\rho}= \int_\Omega F(\sqrt{\alpha} f) d\rho_{D^c}
( f)= \frac{\alpha}{2} \int_\Omega (F^{\prime \prime}(0)f, f) \;
d\rho_{D^c}(f) + \alpha^2 R(\alpha,F,\rho),
\end{equation}
where
$$
R(\alpha,F,\rho)= \int_\Omega g(\alpha,F; f) d\rho_{D^c} (f),
g(\alpha,F; f)= \sum_{n=4}^\infty \frac{\alpha^{n/2-2}}{n!}
 F^{(n)}(0)( f, ..., f).
$$
 We note that
$$ \int_\Omega (F^\prime(0),  f) d\rho_{D^c}(
f)=0,\; \;\; \; \int_\Omega F^{\prime\prime\prime}(0)(f,f,f)
d\rho_{D^c}( f)=0,
$$
because the mean value of $\rho$  (and, hence, of $\rho_{D^c})$ is
equal to zero. Since $\rho\in S_{G}^\alpha(\Omega),$ we have that
the real trace $\rm{Tr} \; D = 1.$ Hence, even the complex trace
$\rm{Tr} \; D^c = 1.$ \footnote{The change of variables  can be
considered as scaling of the magnitude of statistical  (Gaussian)
fluctuations. Negligibly small random  fluctuations $\sigma (\rho)=
\sqrt{\alpha}$ (where $\alpha$ is a small parameter) are considered
in the new scale as standard normal fluctuations. If we use the
language of probability theory and consider a Gaussian random
variables $\xi(\lambda),$ then the transformation (\ref{LHT}) is
nothing else than the standard normalization of this random variable
(which is used, for example, in the central limit theorem):
$\eta(\lambda)= \frac{\xi(\lambda) - E \xi}{\sqrt{E(\xi(\lambda) - E
\xi)^2}}$ (in our case $E \xi=0).$}

We now estimate the rest term $R(\alpha,F,\rho).$ By using
exponential growth of the functional $F(f)$ on the complexification
$\Omega^{{\bf C}}$ of the phase space $\Omega$ we obtain:
 we have for $\alpha \leq 1:$
$$
\vert g(\alpha,F; f)\vert  = \sum_{n=4}^\infty  \frac{\Vert F^{(n)}
(0) \Vert \Vert f \Vert_2^n}{n!} \leq c_F \sum_{n=4}^\infty
\frac{r_F^n  \Vert f \Vert_2^n}{n!}= C_F e^{r_F \Vert f \Vert_2}.
$$
Thus: $ \vert R(\alpha,F,\rho)\vert \leq c_F\int_\Omega  e^{r_F
\Vert f \Vert_2}d\rho_{D^c} (f). $ We obtain:
\begin{equation}
\label{ANN2} <F>_\rho=  \frac{\alpha}{2} \int_\Omega (F^{\prime
\prime}(0)f, f) \; d\rho_{D^c}(f) + O(\alpha^2), \; \alpha \to 0.
\end{equation}
By performing Gaussian integration we finally come the asymptotic
equality (\ref{AI}).

\medskip

{\bf REFERENCES}

\medskip

[1]  Einstein A., Podolsky B., Rosen N., {\it Phys. Rev.}, {\bf 47}
(1935) 777.

Einstein A., {\it The Collected Papers of Albert Einstein}
(Princeton Univ. Press, Princeton) 1993.

[2] Bohr N.,  {\it Phys. Rev.},   {\bf 48} (1935) 696.

[3] Plotnitsky A., {\it Quantum atomicity and quantum information:
Bohr, Heisenberg, and quantum mechanics as an information theory},
in {\it Quantum theory: reconsideration of foundations}, edited by
Khrennikov A. Yu. (V\"axj\"o Univ. Press)  2002, pp. 309-343.

[4] Von Neumann J., {\it Mathematische Grundlagen der
Quantenmechanik} (Springer, Berlin) 1932.

Von Neumann J.,  {\it  Mathematical foundations of quantum
mechanics} (Princeton Univ. Press, Princeton, N.J.) 1955.

[5]  Ballentine L. E.,  {\it Rev. Mod. Phys.}, {\bf 42}   (1970)
358.

[6] Schr\"odinger E.,  {\it Philosophy and the Birth of Quantum
Mechanics,} edited by Bitbol M., Darrigol O. (Editions Frontieres,
Gif-sur-Yvette) 1992; especially the paper of D'Agostino S., {\it
Continuity and completeness in physical theory: Schr\"odinger's
return to the wave interpretation of quantum mechanics in the
1950's}, pp. 339-360.

[7]  Schr\"odinger E., {\it E. Schr\"odinger Gesammelte
Abhandlungen} ( Wieweg and Son, Wien) 1984; especially the paper
{\it What is an elementary particle?}, pp. 456-463.

[8] Bohr N., {\it Niels Bohr: Collected Works,} Vol. {\bf 1-10}
(Elsevier, Amsterdam) 1972-1996.

[9] Plotnitsky A., {\it The Knowable and Unknowable (Modern Science,
Nonclassical Thought, and the ``Two Cultures''} (Univ. Michigan
Press) 2002.

[10] Plotnitsky A., {\it Found. Phys.} {\bf 33} (2003) 1649.

[11] Heisenberg W., {\it Physical Principles of Quantum Theory}
(Chicago Univ. Press) 1930.

[12] Khrennikov A. Yu.,  {\it Prequantum classical statistical model
with infinite dimensional phase-space}, {\it J. Phys. A: Math. Gen.}
{\bf 38} (2005) 9051.

[13] Khrennikov A. Yu., {\it Generalizations of quantum mechanics
induced by classical statistical field theory}, {\it Found. Phys.
Letters}, {\bf 18} (2005) 637.

[14] Khrennikov A. Yu., {\it Interpretation of stationary states in
prequantum classical statistical field theory}, {\it Found. Phys.
Letters}, (2005)- accepted for publication.

[15] Khrennikov A. Yu., {\it Quantum mechanics as an asymptotic
projection of statistical mechanics of classical fields: derivation
of Schr\"{o}dinger's, Heisenberg's and von Neumann's equations},\\
http://www.arxiv.org/abs/quant-ph/0511074.

[16] Bogolubov N. N.,  {\it  Quantum Field Theory} (Taylor and
Francis), 1995.

[17] Segal I., {\it Mathematical Foundations of Quantum Field
Theory} (Benjamin, New York) 1964.

[18] Khrennikov A. Yu., {\it J. Phys.A: Math. Gen.}, {\bf 34} (2001)
9965; {\it Il Nuovo Cimento}, {\bf B 117}   (2002) 267; {\it J.
Math. Phys.}, {\bf 43} (2002) 789;  {\it J. Math. Phys.}, {\bf 44}
(2003) 2471; {\it Phys. Lett. A}, {\bf 316}  (2003) 279; {\it
Annalen der Physik}, {\bf 12}  (2003) 575.

[19] Accardi L., {\it ``The probabilistic roots of the quantum
mechanical paradoxes''} in {\em The wave--particle dualism.  A
tribute to Louis de Broglie on his 90th Birthday,} edited by Diner
S., Fargue  D., Lochak G.,  and Selleri F. (D. Reidel Publ. Company,
Dordrecht) 1984, pp. 297--330.

Accardi L., {\it Urne e Camaleoni: Dialogo sulla realta, le leggi
del caso e la teoria quantistica} (Il Saggiatore, Rome) 1997.

[20] Ballentine L. E., {\it Quantum mechanics} (Englewood Cliffs,
New Jersey) 1989.

[21] Beltrametti E. G., {\it The Logic of Quantum Mechanics}
(Addison-Wesley) 1981.

[22] De Muynck W. M., {\it Foundations of Quantum Mechanics, an
Empiricists Approach} (Kluwer, Dordrecht) 2002.

[23] Gudder S. P., {\it Axiomatic Quantum Mechanics and Generalized
Probability Theory} (Academic Press, New York) 1970.

Gudder  S. P., {\it An approach to quantum probability} in {\it
Foundations of Probability and Physics,} edited by  A. Yu.
Khrennikov, {\it Quantum Prob. White Noise Anal.}, {\bf 13} ( WSP,
Singapore) 2001, pp. 147-160.

[24] Berezin F. A., {\it Method of Second Quantization} (Academic
Press) 1966.

[25] Khrennikov  A. Yu., {\it Equations with infinite-dimensional
pseudo-differential operators,} Dissertation for the degree of
candidate of phys-math. sc., Dept. Mechanics-Mathematics (Moscow
State University, Moscow) 1983.

[26] Khrennikov A.Yu., {\it Supernalysis} (Nauka, Fizmatlit, Moscow)
1997 (in Russian). English translation: (Kluwer, Dordreht) 1999.

[27]  Bach A.,  {\it J. Math. Phys.}, {\bf 14} (1981) 125; {\bf 21}
(1980) 789; {\it Phys. Lett. A},  {\bf 73}  (1979)  287.

[28] De la Pena  L. and  Cetto A. M., {\it The Quantum Dice: An
Introduction to Stochastic Electrodynamics} ( Kluwer, Dordrecht)
1996; Boyer T. H., {\it A Brief Survey of Stochastic
Electrodynamics} in Foundations of Radiation Theory and Quantum
Electrodynamics, edited by Barut A. O. (Plenum, New York) 1980;
Boyer T. H., Timothy H., {\it Scientific American}, pp. 70-78, Aug
1985; see also an extended discussion on vacuum fluctuations in: M.
Scully O., Zubairy M. S., {\it Quantum Optics} (Cambridge University
Press, Cambridge) 1997; Louisell W. H., {\it Quantum Statistical
Properties of Radiation}  (J. Wiley, New York) 1973; Mandel L.  and
Wolf E., {\it Optical Coherence and Quantum Optics} (Cambridge
University Press, Cambridge) 1995.

[29] Cavalleri G., {\it Nuovo Cimento} B, {\bf 112}  (1997) 1193.

[30] Zecca A.  and  Cavalleri G., {\it Nuovo Cimento} B, {\bf 112},
(1997) 1; Cavalleri G.  and Tonni E., "Discriminating between QM and
SED with spin", in C. Carola and A. Rossi, {\it The Foundations of
Quantum Mechanics (Hystorical Analysis and Open Questions)} (World
Sceintific Publ., Singapore), p.111, 2000.

[31] Nelson  E., {\it Quantum fluctuation} (Princeton Univ. Press,
Princeton) 1985.

[32] Albeverio S.,  H\"oegh-Krohn R., {\it  Zeitschrift f\"ur
Wahrscheinlichketstheorie und verwandte Gebite},  {\bf 40} (1977)
59.

[33] Davidson M.,  {\it J. Math. Phys.} {\bf 20}  (1979) 1865; {\it
Physica} A, {\bf  96} (1979) 465.

[34] Morgan P., {\it Phys. Lett.}  A, {\bf  338} (2005) 8; {\bf 321}
(2004) 216.

\end{document}